\documentclass[twocolumn,final,10pt]{elsarticle}

\usepackage{framed,multirow}

\usepackage[english]{babel}
\usepackage[utf8x]{inputenc}
\usepackage[T1]{fontenc}
\usepackage{bm}
\usepackage{amssymb}
\usepackage{xr}

\usepackage{amsmath}
\usepackage{empheq}
\usepackage{graphicx}
\usepackage{epstopdf}
\usepackage{amsfonts}
\usepackage{algorithmic}
\usepackage{textcomp}
\usepackage{booktabs}

\usepackage{algorithm}
\usepackage[colorlinks=false,linkcolor=.,hypertexnames=false,hidelinks]{hyperref}

\usepackage{todonotes}
\usepackage{subcaption}

\usepackage{amsmath,amssymb,amsfonts}
\usepackage{algorithmic}

\usepackage{graphicx}
\usepackage{textcomp}

\usepackage{siunitx}

\usepackage[export]{adjustbox}
\usepackage{latex/medima}

\usepackage{titlesec}

\usepackage{hyperref}

\definecolor{newcolor}{rgb}{.8,.349,.1}

\journal{Medical Image Analysis}

\begin{document}
\newcommand{\bv}[1]{\mathbf{#1}}
\newcommand{\Grad}{\mbox{\boldmath $\nabla$}}
\newcommand{\Div}{\Grad\!\cdot}
\newcommand{\Lap}{\mbox{\boldmath $\Delta$}}
\newcommand{\Curl}{\Grad\!\times}
\newcommand{\delsq}{\nabla^2}
\newcommand{\mean}[1]{\left\langle{#1}\right\rangle}
\newcommand{\infd}{\textrm{d}}
\newcommand{\diff}[2]{\frac{\infd #1}{\infd #2}}
\newcommand{\difftwo}[2]{\frac{\infd^2 {#1}}{\infd {#2}^2}}
\newcommand{\pdiff}[2]{\frac{\partial #1}{\partial #2}}
\newcommand{\pdiffcon}[3]{\left(\frac{\partial #1}{\partial #2}\right)_{#3}}
\newcommand{\pdifftwo}[2]{\frac{\partial^2 {#1}}{\partial {#2}^2}}
\newcommand{\pdifftwomix}[3]{\frac{\partial^2 {#1}}{\partial {#2}\,\partial {#3}}}
\newcommand{\vel}{\bv{v}}
\newcommand{\expu}[1]{\textrm{e}^{#1}}
\newcommand{\ehat}{\bv{e}}
\newcommand{\intomega}{ \mathop{}\!\int\limits_{\Omega}}
\newcommand{\intelement}{ \mathop{}\!\int\limits_{e_k}}
\newcommand{\intelementgen}{ \mathop{}\!\int\limits_{e_k^{xy}}}
\newcommand{\intelementref}{ \mathop{}\!\int\limits_{e_k^{\xi \eta}}}
\newcommand{\be}{\begin{equation}}
\newcommand{\ee}{\end{equation}}
\newcommand{\baa}{\begin{alignat}{2}}
\newcommand{\eaa}{\end{alignat}}
\newcommand\smathcal[1]{
  \mathchoice
    {{\scriptstyle\mathcal{#1}}}
    {{\scriptstyle\mathcal{#1}}}
    {{\scriptscriptstyle\mathcal{#1}}}
    {\scalebox{.7}{$\scriptscriptstyle\mathcal{#1}$}}
  }
\newcommand\sbmathcal[1]{
  \mathchoice
    {{\scriptstyle{\bm{\mathcal{#1}}}}}
    {{\scriptstyle{\bm{\mathcal{#1}}}}}
    {{\scriptscriptstyle{\bm{\mathcal{#1}}}}}
    {\scalebox{.7}{$\scriptscriptstyle{\bm{\mathcal{#1}}}$}}
  }

\newcommand{\note}[2]{ {\bf #1: }#2} 

\renewcommand{\sectionautorefname}{Section}
\let\subsectionautorefname\sectionautorefname
\let\subsubsectionautorefname\sectionautorefname
\let\paragraphautorefname\sectionautorefname

\definecolor{lightgray}{rgb}{0.925,0.925,0.925}

\def\equationautorefname#1#2\null{%
  Eq.#1(#2\null)%
}

\newcommand*{\vertbar}{\rule[-1ex]{0.5pt}{2.5ex}}
\newcommand*{\horzbar}{\rule[.5ex]{2.5ex}{0.5pt}}
\newcommand{\refappendix}[1]{\hyperref[#1]{Appendix~\ref*{#1}}}
\newcommand{\refsupinf}[1]{\hyperref[#1]{Supporting Information \ref*{#1}}}
\newcommand{\refundef}{{\color{red}[x]}}

\newcommand{\beginsupplement}{%
    \setcounter{figure}{0}
    \setcounter{equation}{0}
    \renewcommand{\theequation}{S\arabic{equation}}

    \crefalias{section}{SuppInfoSection}
    \setcounter{section}{0}
}

\newcommand{\setsuppvideo}{%
    \renewcommand*{\figurename}{\hspace{-.3em}}
    \renewcommand*{\thefigure}{Supporting Information Video S\arabic{figure}}
}

\newcommand{\setsuppfig}{%
    \renewcommand*{\figurename}{\hspace{-0.3em}}
    \renewcommand*{\thefigure}{Supporting Information Figure S\arabic{figure}}
}

\newcommand{\colorred}[1]{{\color{red}{#1}}}

\renewcommand{\listfigurename}{List of figures}

\graphicspath{{./figures/}}
\verso{Niek R.F. Huttinga \textit{et~al.}}

\begin{frontmatter}

\title{Gaussian Processes for real-time 3D motion and uncertainty estimation during MR-guided radiotherapy}

\author[1,2]{Niek R.F. Huttinga\corref{cor1}}
\cortext[cor1]{Corresponding author, 
  email: n.r.f.huttinga@umcutrecht.nl}
\author[1,2]{Tom Bruijnen}
\author[1,2]{Cornelis A.T. van den Berg}
\author[1,2]{Alessandro Sbrizzi}

\address[1]{Department of Radiotherapy, Division of Imaging \& Oncology, University Medical Center Utrecht, The Netherlands}
\address[2]{Computational Imaging Group for MR diagnostics \& therapy, Center for Image Sciences, University Medical Center Utrecht, The Netherlands}




\begin{abstract}Respiratory motion during radiotherapy causes uncertainty in the tumor's location, which is typically addressed by an increased radiation area and a decreased dose. As a result, the treatments' efficacy is reduced. The recently proposed hybrid MR-linac scanner holds the promise to efficiently deal with such respiratory motion through real-time adaptive MR-guided radiotherapy (MRgRT). For MRgRT, motion-fields should be estimated from MR-data and the radiotherapy plan should be adapted in real-time according to the estimated motion-fields. All of this should be performed with a total latency of maximally 200 ms, including data acquisition and reconstruction. A measure of confidence in such estimated motion-fields is highly desirable, for instance to ensure the patient's safety in case of unexpected and undesirable motion. In this work, we propose a framework based on Gaussian Processes to infer 3D motion-fields and uncertainty maps in real-time from only three readouts of MR-data. We demonstrated an inference frame rate up to 69 Hz including data acquisition and reconstruction, thereby exploiting the limited amount of required MR-data. Additionally, we designed a rejection criterion based on the motion-field uncertainty maps to demonstrate the framework's potential for quality assurance. The framework was validated in silico and in vivo on healthy volunteer data ($n=5$) acquired using an MR-linac, thereby taking into account different breathing patterns and controlled bulk motion. Results indicate end-point-errors with a 75$^\textrm{th}$ percentile below 1 mm in silico, and a correct detection of erroneous motion estimates with the rejection criterion. Altogether, the results show the potential of the framework for application in real-time MR-guided radiotherapy with an MR-linac. 
\end{abstract}


\end{frontmatter}

\section{Introduction}

Motion during abdominal radiotherapy decreases the efficacy of treatments due to an uncertain tumor location. This uncertainty can be reduced in several ways, one of which is to estimate the tumor's motion from MR-data acquired during radiation with an MR-linac \citep{Lagendijk2008,raaymakers2009integrating,Keall2014,Mutic2014}, a hybrid device which combines an MR-scanner and a radiotherapy LINAC.
The ultimate goal with the MR-linac is real-time adaptive MR-guided radiotherapy (MRgRT). This requires a continuous loop comprising tumor motion estimation, followed by corresponding radiation beam adjustments. MR-guided radiotherapy introduces several major technical challenges, one of which is to reconstruct accurate 3D motion-fields in real-time, from a stream of MR-data. The required speed for this reconstruction is determined by the expected velocities; for slowly-moving tumors such as prostate tumors 1 Hz could be sufficient, but tumors subject to respiratory motion requires at least 5 Hz \citep{Murphy2002,keall2006management}. 

Here, we focus on the latter category, which requires MR-data acquisition and motion reconstruction with a frame rate of at least 5 Hz. Motion-fields can be estimated from MR-images by means of image registration (image-based), or directly from $k$-space data ($k$-space-based) \citep{huttinga2020mr}. Reconstructing 3D motion-fields of the abdomen and thorax at this rate is currently still challenging. To increase the achievable frame rate of 3D motion reconstructions, prior assumptions of images and/or motion-fields need to be included in the reconstruction. A frequently used strategy is to exploit an a priori built model with a two-step approach: 1) a calibration or training phase to build a patient-specific motion or image model prior to radiation; 2) a real-time phase during treatment, in which 3D information is reconstructed from a minimal amount of rapidly acquired MR-data by exploiting the model from the training phase. 
Examples of image-based methods proposed for radiotherapy include MR-SIGMA by \citet{Feng2020}, and approaches based on interleaved orthogonal cine images \citep{stemkens2016image,Bjerre2013,Paganelli2018,Mickevicius2017}. An example of a $k$-space-based method for radiotherapy is the authors' real-time low-rank MR-MOTUS \citep{Huttinga2021,huttinga2020mr,huttinga2021realtime}. MR-SIGMA estimated 3D MR-images at 3.3 Hz, \citet{stemkens2016image} achieved 3D motion-field reconstructions at about 2 Hz using image registration and cine-MRI, and MR-MOTUS achieved 3D motion-field reconstructions directly from $k$-space data at 6.7 Hz. Although considerably different in the modelling aspect, all the methods mentioned above employed a two-step approach.

It should be noted, however, that this two-step approach relies on the assumption that the motion in the training and real-time phases are similar. Although this is likely true in most cases, several practical scenarios such as bulk motion, or a change in breathing pattern could reduce the validity of this assumption. This could therefore result in erroneous motion estimates, which - if left undetected - could eventually lead to harmful radiation to the patient. To warrant the patient's safety in such scenarios, methods for real-time MRgRT should therefore ideally not only estimate motion in real-time, but should also provide some measure of reliability. In a practical setting, this could be used for real-time quality assurance during radiotherapy, e.g. to halt and resume the radiation treatment according to the degree of confidence.

In this work, we present a probabilistic framework to simultaneously quantify 3D motion and provide a measure of reliability in real-time, which thereby addresses two critical needs for real-time MRgRT. This work is an extension of the preliminary work presented as conference abstracts in \citet{sbrizzi2019acquisition} and \citet{huttinga2021joint}. Our framework is based on the previously discussed two-step reconstruction approach. Firstly, in the training phase, a model for respiratory-resolved motion-fields is built that allows to represent 3D motion-fields with few coefficients. Secondly, in the inference phase, these representation coefficients are estimated from three mutually orthogonal readouts of MR-data in real-time, thereby exploiting the motion model built in the training phase. 

The idea to extract motion information directly from few readouts was motivated by the success of the authors' MR-MOTUS method \citep{huttinga2021realtime}, which reconstructs low-dimensional motion-field representation coefficients from few readouts of $k$-space data. The idea to use three mutually orthogonal readouts in order to do this is motivated as follows. A single readout of $k$-space data that crosses the $k$-space center effectively contains a projection of the excited FOV in the readout direction. This can be seen by transforming the readout to image space. It therefore mostly contains information of motion in the direction of the readout. Consequently, a set of three mutually orthogonal readouts contains information of motion in all directions. Based on the two observations above, we hypothesized that the low-dimensional motion-field representation coefficients can directly be inferred from the three mutually orthogonal readouts. That is, we assume the motion-field representation coefficients are a function of the three readouts. 

Here, we propose to learn this underlying function via a probabilistic machine learning regression technique called Gaussian Processes (GP). A GP requires a calibration phase to tune its internal parameters based on a small training set, which takes $\approx 0.5$ seconds in this work. The trained GP can then be used for real-time inference of the posterior distribution of the 3D motion-field representation coefficients, given the three mutually orthogonal readouts. This step exploits the availability of a closed-form analytical expression for the posterior of a GP, which allows for sub-millisecond computations ($\approx 0.1$ milliseconds per dynamic in this work). Combining the resulting motion-field representation coefficients with the motion model, this eventually allowed for 3D motion-field reconstruction at frame rates as high as 69 Hz. Moreover, the posterior distribution as inferred by the GP captures not only the most likely motion-field estimate corresponding to the input data, but also the corresponding estimation uncertainty. The latter provides a measure of both the model-related and measurement-related uncertainties (respectively the epistemic and aleatoric uncertainties). As a consequence, a measure of motion model reliability is provided. We therefore hypothesize that the GP posterior uncertainty can be used for real-time quality assurance, i.e. to detect potentially erroneous motion estimates of the proposed framework.

Altogether, the framework could perform simultaneous real-time 3D motion-field estimation with real-time quality assurance. We assessed the accuracy of the presented framework based on several in silico and in vivo tests, and tested our hypothesis regarding the value of the GP posterior uncertainty for real-time quality assurance. For the latter, we designed a rejection criterion based on the uncertainty that flags dynamics with potential erroneous motion estimates in real-time. Both the motion estimation and rejection criterion were evaluated on data simulated using a digital XCAT phantom and in vivo data of 5 healthy volunteers acquired on an MR-linac, thereby considering different types of breathing and bulk motion.

\section{Theory}
\subsection{A general introduction to Gaussian Processes}
A Gaussian Process (GP) \citep{rasmussen2003gaussian} models a Gaussian probability distribution over functions, and can therefore be considered as an extension of the multi-variate Gaussian distribution, which models a Gaussian probability distribution over vectors. Gaussian Processes are frequently applied to regression problems by assuming the following model for noisy measurements $\bv{y}_t
\in\mathcal{Y}\subset\mathbb{R}^{N_\mathcal{Y}}$ at samples $\bv{x}_t \in \mathcal{X} \subset \mathbb{C}^{N_\mathcal{X}}$: 
\begin{equation}\label{eq:noisymeasurementmodel} \bv{y}_t = \bm{y}(\bv{x}_t)+\bm{\epsilon},\end{equation}
where $\bm{\epsilon}\in\mathcal{X},\bm{\epsilon}\sim\mathcal{N}(\bv{0},\sigma_n^2 \bv{I})$, and $\bm{y}:\mathcal{X}\rightarrow\mathcal{Y}$ is the underlying process. In case a function is drawn from a GP, and evaluated at a finite collection of $N_\mathcal{T}$ samples, $\bv{x}_\mathcal{T}:=[\bv{x}_1,\dots,\bv{x}_{N_\mathcal{T}}]\in\mathbb{R}^{N_\mathcal{X} \cdot N_\mathcal{T}}$, the vertically concatenated corresponding function values $\bv{y}_\mathcal{T}:=[\bv{y}_1;\dots; \bv{y}_{N_\mathcal{T}}]\in\mathbb{R}^{N_\mathcal{Y} \cdot N_\mathcal{T}}$ follow a multi-variate Gaussian distribution. A GP is completely characterized by a mean {\it function} $m(\bv{x})$, and kernel covariance {\it function} $k(\bv{x}_i,\bv{x}_j)$, which in turn specify the mean vector and covariance matrix of the corresponding multi-variate Gaussian distribution: \begin{equation}
\label{eq:drawfromgp}
\bv{y}_\mathcal{T}|\bv{x}_\mathcal{T} \sim \mathcal{N}(\bv{m}_\mathcal{T},\bv{K}_{\mathcal{T},\mathcal{T}}).
\end{equation}
Here, $\mathcal{T}$ is defined as the training set $\mathcal{T}:=\{(\bv{x}_i,\bv{y}_i)\}_{i=1}^{N_\mathcal{T}}$, and $\bv{m}_\mathcal{T}$ and $\bv{K}_{\mathcal{T},\mathcal{T}}$ denote the mean vector and covariance matrix, which are computed by evaluating respectively the mean and kernel function for all training samples in $\mathcal{T}$. 

The kernel function characterizes the properties of the underlying process (in this case $y$), such as smoothness or periodicity. Without loss of generality, a zero-mean GP is typically assumed with $m \equiv 0$, in which case the GP is completely characterized by the kernel function $k(\bv{x}_i,\bv{x}_j)$. The kernel function is typically a function of the distance between its two inputs, and is parameterized by hyperparameters $\bm{\theta}$ that determine its form: $k(\bv{x}_i,\bv{x}_j):=f(\lVert \bv{x}_i-\bv{x}_j\rVert_2 \ |\bm{\theta})$.  

Evidently, correct tuning of $\bm{\theta}$ directly influences the GP's regression performance. Therefore, two steps are followed to perform GP-based regression. In the first step, the GP hyperparameters $\bm{\theta}$ are estimated by maximum likelihood estimation (MLE) on the joint Gaussian likelihood \eqref{eq:drawfromgp}. The result is a fully-determined GP kernel function, specifying the properties of the functions that best fit the training measurements. In the second step, a posterior distribution over the function values $y_\mathcal{Q}$ at samples $\mathcal{Q}$ is computed by conditioning on the training data $\mathcal{T}$. This yields the Gaussian posterior distribution (\citet{rasmussen2003gaussian}, Eq. [A.6]):
\begin{equation}
\label{eq:gpposterior_general}
\bv{y}_\mathcal{Q} | \mathcal{T}  \sim \mathcal{N}\left(\bv{m}_\mathcal{Q} , \bm{\Sigma}_\mathcal{Q} \right),
\end{equation}
with 
\begin{alignat}{2}
\bv{m}_\mathcal{Q} &= \bv{K}_{\mathcal{T},Q}^T\left(\bv{K}_{\mathcal{T},\mathcal{T}}+\sigma_n^2 \bv{I}\right)^{-1} \bv{y}_\mathcal{T}, \label{eq:gpposteriormean}\\
\bm{\Sigma}_\mathcal{Q} &= \bv{K}_{QQ} - \bv{K}_{\mathcal{T},Q}^T \left(\bv{K}_{\mathcal{T},\mathcal{T}}+\sigma_n^2 \bv{I}\right)^{-1} \bv{K}_{\mathcal{T},Q}. \label{eq:gpposteriorcovariance}
\end{alignat}
In particular, the diagonal elements of the covariance matrix $\bm{\Sigma}_\mathcal{Q}$ define the GP prediction uncertainty. 

One of the main strengths of the GP framework is the availability of the closed-form analytical expressions in \autoref{eq:gpposteriormean}-\eqref{eq:gpposteriorcovariance} that completely characterize the posterior distribution in \autoref{eq:gpposterior_general}. The computations involve the matrices $\bv{K}_{\mathcal{T},\mathcal{Q}}\in\mathbb{R}^{N_\mathcal{Y}N_\mathcal{T} \times N_\mathcal{Y} }, \bv{K}_{\mathcal{T},\mathcal{T}}\in\mathbb{R}^{N_\mathcal{Y}N_\mathcal{T}\times N_\mathcal{Y}N_\mathcal{T} }$ and $\bv{K}_{\mathcal{Q},\mathcal{Q}}\in\mathbb{R}^{N_\mathcal{Y}\times N_\mathcal{Y} }$, where $N_\mathcal{T}$ denotes the number of training samples and $N_\mathcal{Y}$ the number outputs.

\begin{figure*}[h!]
    \centering
    \includegraphics[width=.9\textwidth]{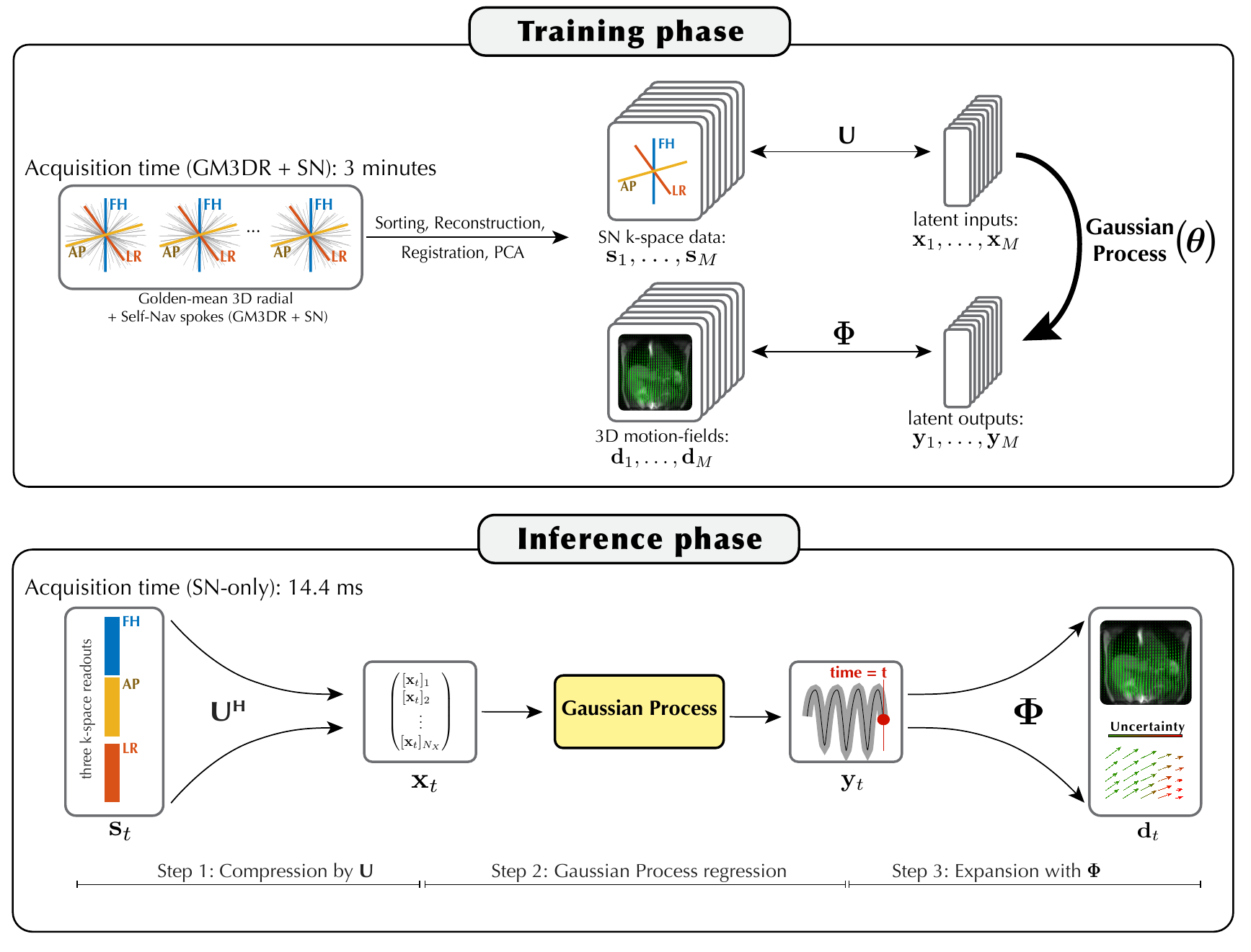}
    \caption{Framework overview. In the training phase, GM3DR+SN data is acquired during 3 minutes of free-breathing. The motion model and spoke compression basis are built with PCA on respectively respiratory-sorted GM3DR+SN and SN spokes. The motion model requires motion-fields, which are obtained by Optical Flow \citep{zachiu2015improved} on respiratory-sorted images. Finally, GP hyperparameters $\bm{\theta}$ are obtained by maximum likelihood estimation on the training set $\mathcal{T}=\{(\bv{x}_i,\bv{y}_i)\}_{i=1}^M$, consisting of the compressed spoke representations $\bv{x}_i$ and motion-fields representation coefficients $\bv{y}_i$. In the inference phase three steps are followed: 1) SN spokes are compressed to $\bv{x}_t = \bv{U}^H \bv{s}_t$; 2) the posterior distribution $P(\bv{y}_t|\mathcal{T})$ is computed with the GP from $\bv{x}_t$; 3) a 3D motion-field $\bv{d}_t$ and a corresponding uncertainty map are inferred by expanding the GP posterior distribution with the motion model $\bm{\Phi}$.}
    \label{fig:frameworkoverview}
\end{figure*}

\subsection{Proposed framework: integrating Gaussian Processes in motion modeling}
The technical challenge for this application is to perform a real-time reconstruction of a complete 3D motion-field and corresponding reconstruction uncertainties from few readouts that can be rapidly acquired. This work considers Gaussian Processes for this purpose for two main reasons. Firstly, for the targeted application the computations in \autoref{eq:gpposteriormean}-\eqref{eq:gpposteriorcovariance} can be performed in the order of milliseconds. Secondly, the probabilistic nature of GPs gives access to reconstruction uncertainties, which will prove valuable for quality assurance. 

We identify 3 practical challenges when trying to apply GPs in our context. Firstly, GPs suffer from the curse of dimensionality, and are therefore challenging to apply to high-dimensional inputs (e.g. $>300$ total samples on the three mutually orthogonal readouts); distances in high-dimensional input-space become uninformative \citep{bengio2006curse}, while these distances - as evaluated by the kernel function - form the basis of the GP theory. 

Secondly, GPs are mostly applied to regression problems with scalar functions, since modeling correlations in outputs is a challenging task and requires a non-trivial extension of the framework \citep{bonilla2007multi}. To overcome these first two challenges, we propose a linear compression of both the input space and the output space, and learn a GP that maps from the latent input space $\mathcal{X}\subset \mathbb{R}^{N_\mathcal{X}}$ to the latent output space $\mathcal{Y}\subset\mathbb{R}^{N_\mathcal{Y}}$. The schematic pipeline from few readouts $\bv{s}_t\in\mathcal{S}\subset\mathbb{C}^{N_\mathcal{S}}$, to motion-fields $\bv{d}_t \in\mathcal{D}\subset \mathbb{R}^{3N}$, then becomes 
\begin{equation}\label{eq:methodpipeline}
\mathcal{S} \xrightarrow[]{\hspace{.2cm}\bv{U}^H\hspace{.2cm}} \mathcal{X} \xrightarrow[]{\hspace{.2cm}\textrm{GP}\hspace{.2cm}} \mathcal{Y} \xrightarrow[]{\hspace{.2cm}{\bm{\Phi}}\hspace{.2cm}} \mathcal{D}, \end{equation} where $\bv{U}$ and $\bm{\Phi}$ denote orthogonal bases of the latent input and output space, and $\bv{U}^H$ the Hermitian transpose of $\bv{U}$. This pipeline is also visually outlined in \autoref{fig:frameworkoverview}. Assuming $N_\mathcal{Y}\ge 1$ uncorrelated dimensions, each element $\left[\bv{y}_t\right]_j$ ($j=1,\dots,N_\mathcal{Y}$) of $\bv{y}_t$ can be modelled with a separate GP, each of which takes $\bv{x}_t$ as input. 

Now that the general pipeline is outlined, the third practical challenge can be identified. Training of the GPs in the pipeline \eqref{eq:methodpipeline} requires training sets $\mathcal{T}_j=\{(\bv{x}_t,\left[\bv{y}_t\right]_j) \}_{t=1,\dots,N_\mathcal{T}}$, for $j=1,\dots,N_\mathcal{Y}$. On the one hand, the GPs are required to constantly evaluate distances between $\bv{x}_t$'s. To do this consistently, all $\bv{x}_t$ should thus measure the same $k$-space locations but at a different time instance, which in turn requires the same $k$-space locations for all $t$. On the other hand, the targets $\bv{y}_t$ will be derived from 3D images, and thus require $k$-space samples from the entire range of $k$-space coordinates. To meet both requirements, we employed an interleaved acquisition of mutually orthogonal self-navigation $k$-space spokes (SN spokes) and golden-mean 3D radial (GM3DR)\cite{johnson2017hybrid}. The SN spokes are acquired along the three mutually orthogonal axes: feet-head (FH), anterior-posterior (AP), and left-right (LR). Before further processing the $k$-space spokes are transformed to image space through an FFT along the readout direction. Next, the vertically concatenated SN spokes at dynamic $t$, denoted as $\bv{s}_t\in\mathcal{S}$ are processed to GP inputs $\bv{x}_t\in\mathcal{X}$, and GM3DR spokes at dynamic $t$ to GP outputs $\bv{y}_t\in\mathcal{Y}$. For the SN spokes, the processing includes the computation of a compressed representation with respect to an orthonormal basis $\bv{U}$. For GM3DR spokes, the processing comprises a respiratory-resolved image reconstruction, followed by an image registration, the construction of an orthogonal motion model $\bm{\Phi}$ via PCA, and finally the computation of a representation with respect to $\bm{\Phi}$. More details on these processing steps will be given in the subsequent sections. The construction of the GP training sets $\bm{\mathcal{T}}_j$ from the acquired SN and GM3DR spokes is also visualized in \autoref{fig:frameworkoverview}.

The inference pipeline \autoref{eq:methodpipeline} requires three main steps: 1) a compression of readout spokes to the latent input space using $\bv{U}$, 2) regression between the latent input and output space by a GP with hyperparameters $\bm{\theta}$, 3) an expansion of the latent output space to 3D motion-fields using a motion model $\bm{\Phi}$. These steps are also visualized at the bottom of \autoref{fig:frameworkoverview}. All three sets of parameters ($\bv{U},\bm{\theta},\bm{\Phi}$) mentioned above are obtained from respiratory-resolved training data, which was in turn acquired by sorting all GM3DR spokes with a 1D respiratory motion surrogate extracted from the closest feet-head spoke\cite{feng2016xd,Huttinga2021}. Details including how the required parameters $\bv{U}, \bm{\theta}$ and $\bm{\Phi}$ in the pipeline \autoref{eq:methodpipeline} are obtained will be discussed next, organized in sections. The first step in the pipeline, i.e. a compression of the acquired space, will be discussed in \autoref{section:inputspacecompression}. The second step, i.e. the GP regression, will be discussed in \autoref{section:gpregressiondetails}. The third step, i.e. the expansion from GP outputs to motion-fields, will be discussed in \autoref{section:outputspacecompression}. Finally, we also investigate the potential of the framework outlined above for quality assurance. To this extent, we designed and evaluated a rejection criterion based on the GP's reconstruction uncertainty. The rejection criterion is evaluated in parallel with the inference pipeline, rejecting unreliable motion estimates due to undesirable motion. Details on this are discussed in \autoref{sec:rejection_criterion}.

\subsubsection{Step 1: Input space compression with \texorpdfstring{$\bv{U}$}{\textbf{U}}}
\label{section:inputspacecompression}
We empirically observed a sensitivity of the reconstruction to cardiac motion, which resulted in high-frequency oscillations on top of the respiratory motion. Since this is undesirable, the $k$-space readouts were first transformed to image space with an FFT along the readout direction, and subsequently samples on the readouts (radial spokes) were thresholded based on their temporal frequency spectrum; any sample with a large contribution from the cardiac frequency range was removed. This thresholding typically resulted in an automatic selection of the liver, and excluded the heart and large arteries. After thresholding, respiratory-sorting was performed on the spokes, and PCA was performed along the respiratory dimension on the respiratory-sorted matrix of spokes $\bv{S}:=[\bv{s}_1,\dots,\bv{s}_{20}]$, such that $\bv{S}\approx\bv{U}\bv{V}^H$. Here, and in the rest of this work, the number of respiratory bins was set to 20. The $\bv{s}_i \ (i=1,\dots,20)$ were obtained by averaging over all SN spokes that ended up in the $i$-th bin. The number of columns in $\bv{U}$ and $\bv{V}$ for the approximation $\bv{S}\approx\bv{U}\bv{V}^H$ was restricted to $\max(4,N_\mathcal{Y}+1)$, where $N_\mathcal{Y}$ is the rank of the motion model as described above. This restricts the compression of the inputs to no less than 4 samples. This was empirically observed to improve the flexibility of the framework to correctly perform the regression. Finally, data compression at time $t$ was performed by $\bv{x}_t = \bv{U}^H \bv{s}_t$.

\subsubsection{Step 2: GP regression between latent input and output space}
\label{section:gpregressiondetails}
For the GPs we chose to use a Mat\'{e}rn kernel with $\nu=3/2$ and automatic relevance determination (ARD). The Mat\'{e}rn class of kernel functions is an extension of the class of radial basis functions, and restricts the functions that can be drawn from the GP to be $\lfloor\nu\rfloor$ times differentiable. With $\nu=3/2$, this thus results in a GP that models once-differentiable functions. In general, kernel functions return the correlation between its two inputs. In case of an ARD kernel, the relevance of each input dimension in this correlation computation is specified by the length-scales $l_1,\dots,l_{N_\mathcal{X}}$. For the Mat\'{e}rn-3/2 kernel, we have 
\begin{multline}
\label{eq:matern32kernel}
k(\bv{x}_i,\bv{x}_j| \sigma_n,l_1,\dots,l_{N_\mathcal{X}})=\\ \sigma_n^2 \left(1+\sqrt{3}\lVert\bv{x}_i-\bv{x}_j\rVert_{\bv{L}} \right) \exp\left(-\sqrt{3}\lVert\bv{x}_i-\bv{x}_j\rVert_{\bv{L}}\right),
\end{multline}
Here, $\bm{\theta}:=\{\sigma_n,l_1,\dots,l_{N_\mathcal{X}}\}$ denote the kernel hyperparameters, and we defined $\bv{L}:=\textrm{diag}\left(1/l_1^2,\dots,1/l_{N_\mathcal{X}}^2\right)$ and the norm of a vector $\bv{a}$ with respect to a positive definite matrix $\bv{B}$ as $\lVert\bv{a}\rVert_{\bv{B}}:=\sqrt{\bv{a}^T\bv{B}\bv{a}}$. Theoretically, the standard deviation $\sigma_n$ of the target noise $\bm{\epsilon}$ in \autoref{eq:noisymeasurementmodel} could also be optimized, but here we manually fixed $\sigma_n$ such that the 95\% confidence interval approximately contained the neighboring bin's training targets (see first column in \autoref{fig:robustnesstest_allvols}). With inputs $\bv{x}_t \in \mathbb{R}^{N_\mathcal{X}}$, this resulted in a total of $N_\mathcal{X}$ optimizable GP hyperparameters. Finally, $N_\mathcal{Y}$ pair-wise training sets were constructed as $\mathcal{T}_j:=\{(\bv{U}^H\bv{s}_t,\left[\bv{y}_t\right]_j)\}_{t=1,\dots,20}$, and $N_\mathcal{Y}$ sets of hyperparameters - one for each GP - were obtained through MLE on \autoref{eq:drawfromgp}. The total time of the training was about 0.5 seconds. For the GP hyperparameter optimization we made use of the Gaussian Processes for Machine Learning (GPML) Matlab toolbox \citep{rasmussen2010gaussian}. Once the training is performed, the resulting GP can infer a posterior distribution $P(\bv{y}_t | \bv{x}_t)$ of the motion-field representation coefficients $\bv{y}_t$, given the compressed representation of the spokes $\bv{x}_t=\bv{U}^H\bv{s}_t$. This inference process took about $0.1$ milliseconds per dynamic.

\subsubsection{Step 3: Expansion from latent output space to motion-fields}
\label{section:outputspacecompression}
To expand the posterior distribution on $\bv{y}_t$ to motion-fields, a motion model is required. To derive the motion model $\bm{\Phi}$, first respiratory-resolved MR-images $\bv{I}:=[\bv{I}_{1},\dots,\bv{I}_{20}]\in\mathbb{C}^{N\times 20}$, were reconstructed with an L1-ESPIRiT reconstruction without temporal regularization with the BART toolbox \citep{uecker2015berkeley}, where $N$ denotes the number of voxels. Subsequently, respiratory-resolved motion-fields were obtained by image registration \citep{zachiu2015improved} between the end-exhale dynamic and all other dynamics, and stored in a matrix, $\bv{D}=[\bv{d}_1,\dots,\bv{d}_{20}] \in \mathbb{R}^{3N \times 20}$. Similarly as in the two-phase motion-field reconstruction approaches, PCA was performed along the respiratory dimension of $\bv{D}$ to obtain a linear motion model $\bm{\Phi}$. Typically, most information of $\bv{D}$ is concentrated in its first principal components \citep{zhang2007patient, stemkens2016image, li2011pca,Huttinga2021}, allowing to truncate the number of columns of $\bm{\Phi}$ from 20 to $N_\mathcal{Y} \ll 20$, without sacrificing too much representational power. Thus, $\bm{\Phi}\in\mathbb{R}^{3N \times N_\mathcal{Y}}$, and $\bv{D}$ is approximately low-rank, with rank $N_\mathcal{Y}$. In all experiments in this work, we normalized all columns in $\bm{\Phi}$ to unit norm, and determined the rank via the L-curve of the explained variance of $\bv{D}$ as a function of the number of components. This resulted in $1\le N_\mathcal{Y}\le 3$ for all performed tests, and a motion model which approximates high-dimensional motion fields $\bv{d}_t\in\mathbb{R}^{3N\times1}$ with low-dimensional representation coefficients $\bv{y}_t \in \mathbb{R}^{N_\mathcal{Y}}$:  $\bv{d}_t \approx \bm{\Phi}\bv{y}_t.$

A posterior distribution of the motion-field at time $t$ can be derived by applying the motion model to the posterior distribution of the representation coefficients - $P(\bv{y}_t | \bv{s}_t):=\mathcal{N}(\boldsymbol{\mu}_t,\boldsymbol{\Sigma}_t)$ - as output by the GP. Using basic properties of the multivariate Gaussian distribution, we can derive
\begin{equation}
    \label{eq:gp_dvf_distribution}
    \bv{d}_t := \bm{\Phi} \bv{y}_t \sim \mathcal{N}\left(\bm{\Phi}\bm{\mu}_t , \bm{\Phi}\bm{\Sigma}_t \bm{\Phi}^T \right).
\end{equation}
The spatio-temporal uncertainties $\bv{P}_t$ in the motion-fields can be obtained as the diagonal of the covariance matrix in \autoref{eq:gp_dvf_distribution}, i.e.  $\bv{P}_t:=\textrm{diag}(\bm{\Phi}\bm{\Sigma}_t \bm{\Phi}^T)$. Note that this uncertainty map should be treated with caution. For the derivation above it is assumed that $\bm{\Phi}$ is valid, and that there no uncertainty associated with it. Hence, the spatio-temporal uncertainties in the motion-fields $\bv{d}_t$ can only be reasonably interpreted when this assumption is true. Unlike the uncertainty on $\bv{d}_t$, the uncertainty in $y_t$ has a useful meaning also when $\bm{\Phi}$ is no longer valid. This forms the basis of the rejection criterion, as will be discussed next.

\subsection{Online quality assurance via posterior uncertainty}
\label{sec:rejection_criterion}
We propose a rejection criterion based on the posterior uncertainty, which flags dynamics with uncertain predictions as follows. A new measurement can be considered unreliable when the corresponding prediction uncertainty is substantially higher than those evaluated on the training set. To this extent, we defined the rejection threshold $\tau$ such that
\begin{equation}\label{eq:rejthreshold} P( \sigma_t < \tau | \ t \in \mathcal{T} )=\alpha,\end{equation}
where $\sigma_t$ is a measure of uncertainty. Given this definition of $\tau$, any prediction with an uncertainty exceeding $\tau$ will be rejected. Effectively, this ensures a false positive rejection rate - i.e. the chance of incorrect rejections - of approximately 1-$\alpha$. Note that $\tau$ in \autoref{eq:rejthreshold} is equivalent to the $\alpha$-th percentile over all $\sigma_t$ evaluated on the training set data. In all experiments in this work we set $\alpha=0.95$. Each of the $N_\mathcal{Y}$ GPs outputs a measure of uncertainty (i.e. the variance), which are combined to a single measure of uncertainty in order to evaluate the rejection criterion. To this end, we took a weighted linear combination, where the weights were determined by the singular values in the motion model. This leads to $\sigma_t = \sum_{j=1}^{N_\mathcal{Y}} s_j^2 [\Sigma_t]_{j,j}$, where $s_j$ is the singular value corresponding to the $j$-th singular vector of the motion model $\bm{\Phi}$.

\section{Methods}

\subsection{Data simulation and acquisition}
\label{section:details_data}
For the numerical simulations, data were generated using the digital XCAT phantom \citep{segars20104d} with manually added MR-contrast. XCAT requires two inputs, an abdominal and chest waveform, and returns dynamic volumetric images with respiratory motion (cardiac motion was disabled) and corresponding ground-truth motion-fields. To simulate realistic motion, $\cos^4$ waveforms were used, hysteresis was simulated through a phase-delay between the two input waveforms, and the variations in the extreme respiratory positions were randomly generated as 1\% and 2\% of the waveform amplitudes for respectively end-exhale and end-inhale. All simulated motion-fields were post-processed with cid-X \citep{Eiben2020} to ensure realistic and invertible motion-fields. A complex reference volume in end-exhale was generated by adding a smoothly varying phase, and the resulting motion-fields were applied to this reference volume by cubic interpolation, resulting in a series of dynamics. Multi-channel images were simulated by multiplying each dynamic with 8 (static) coil sensitivity maps simulated with BART \citep{uecker2015berkeley}. Finally, 800 readouts of $k$-space data were simulated for each dynamic with a NUFFT \citep{barnett2018parallel} evaluated on the GM3DR+SN trajectory, and complex noise was added to the resulting $k$-space data to achieve an SNR of 60. In total four types of breathing modes were simulated for our experiments: normal, chest-only, abdominal-only, and amplitude drifts. Chest-only and abdominal-only breathing were simulated by setting respectively the abdominal and chest waveforms to zero, amplitude drifts were simulated by adding a linearly increasing shift to the normal breathing waveforms. For each breathing mode, approximately five breathing cycles were generated, resulting in a total of 100 dynamics. The total data per breathing mode corresponds to an acquisition of approximately 6 minutes. As described later in \autoref{section:insilicoexperiments}, half of the normal breathing phase data were used for training, and the other half for testing. Hence, the in silico training was performed on data that could in practice be acquired in 3 minutes. This is similar to the in vivo experiments, as will be discussed later in \autoref{section:invivoexperiments}.

For the in vivo acquisitions, data were acquired on a 1.5T Elekta Unity MR-linac (Elekta AB, Stockholm, Sweden) from healthy volunteers during free-breathing. All experiments were approved by the institutional review board, carried out in accordance with the relevant guidelines and regulations, and written informed consent was obtained from all volunteers prior to the experiments. In all experiments, we used the standard clinical 8-element radiolucent array with anterior and posterior coils and a steady-state spoiled gradient echo sequence (SPGR) with TR = 4.8 ms, TE = 1.8 ms, FA = 20\textdegree, FOV = 30 cm x 30 cm x 30 cm, BW = 540 Hz, resolution = 3 mm x 3 mm x 3 mm. We considered two modes for the data acquisition. In the first mode, data were acquired with a golden-mean 3D radial (GM3DR) kooshball trajectory \citep{chan2009temporal}, interleaved every 31 spokes with three self-navigation spokes (SN-spokes); along feet-head (FH), anterior-posterior (AP), and left-right (LR). This first mode is referred to as the GM3D+SN mode. Since the proposed framework can infer a 3D motion-field from only a single set of SN-spokes, this interleaved acquisition allows for inference at $1000/(31\times 4.8)\approx 6.7$ Hz. The second mode served to test the feasibility of high-speed inference. In this mode only the SN-spokes were acquired, which allows for inference at $1000/(3\times 4.8)\approx 69$ Hz. The second mode is referred to as the SN-only mode.

\subsection{In silico experiments}
\subsubsection{End-point-error analysis and model applicability tests}
\label{section:insilicoexperiments}
To validate our framework, we performed an in silico end-point-error (EPE) analysis. These EPEs were computed per voxel, as the magnitude of the difference between the reconstructed $\hat{\bv{d}}_t$ and ground-truth motion-fields $\bv{d}_t$: 
\begin{equation}
    \label{eq:epe}
    \textrm{EPE}([\hat{\bv{d}}_t]_i,[\bv{d}_t]_i):=\left\lVert [\hat{\bv{d}}_t]_i-[\bv{d}_t]_i \right\rVert_2,
\end{equation}
where $[\cdot]_i$ denotes the $i$-th voxel. We considered the scenario where we train only on data simulated during normal breathing, and performed inference on all types of breathing: normal, chest-only and abdominal-only breathing, and amplitude drifts. The normal breathing data were separated in two sets, a train and a test set. We analyzed the results in two ways. Firstly, we compared boxplots of the EPEs of a model trained for normal breathing, and applied to all breathing patterns. We hereby separated the EPEs statistics in dynamics rejected by our framework and the non-rejected dynamics, which allowed to see the effect of the proposed rejection criterion on the EPEs. Secondly, we analyzed the correlation between the GP posterior uncertainty and EPEs.

\subsection{In vivo experiments}
\label{section:invivoexperiments}
\subsubsection{In vivo robustness test}
\label{section:experiments_robustnesstest}
To assess the robustness of the proposed framework we trained the model on three minutes of free-breathing and performed inference over all remaining GM3DR spokes. Data were simulated with the digital XCAT phantom, and acquired from 5 volunteers on the MR-linac, as described in \autoref{section:details_data}. The volunteers were not instructed regarding their breathing. The preparation phase was performed on three minutes of data, as described in \autoref{section:inputspacecompression}, \autoref{section:gpregressiondetails} and \autoref{section:outputspacecompression}. Inference was performed on the remainder of the data according to the pipeline in \autoref{eq:methodpipeline}, and finally the rejection criterion was evaluated as described in \autoref{sec:rejection_criterion}. 

The observable motion in FH spokes is typically used as a surrogate for respiratory motion \citep{feng2016xd}. The first principal component of the motion model most significantly contributes to the final reconstructed motion-field. Hence, the motion in the FH spokes should be similar to the motion in the first principal component of the motion model, i.e. the output of the first GP. To this extent, we qualitatively validated the output of the first GP (posterior mean) by comparison with the projection images obtained from these FH $k$-space spokes. These projections were obtained by performing an FFT along the readout direction of the spokes. The comparison was performed on data acquired about two minutes after the training data.

\subsubsection{Feasibility tests of high-speed inference at 69 Hz}
\label{section:highspeedfeasibility_experiments}
In order to perform the input data dimension reduction, train the GP, and build a motion model, the presented framework requires both golden-mean 3D radial kooshball and SN spokes in the training phase. In the inference phase, however, the golden-mean radial spokes are not required, and the use of only the feet-head spokes would lead to a much higher inference frequency. With this experiment we investigate the feasibility of such high-speed inference. 

To this extent, data were acquired in two phases. In the first phase, GM3DR+SN data were acquired in order to train the model. In the second phase, only the SN spokes were acquired at 69 Hz. Data were continuously acquired from four healthy volunteers (volunteers 2-5) over 9 minutes, with the same acquisition parameters as described in \autoref{section:details_data}. The first 7 minutes of data were acquired with GM3DR+SN, and the last 2 minutes with SN-only. To mimic a realistic setting, only the first three minutes of the GM3DR+SN were used to perform all preparation steps in the training phase. The rejection criterion was calibrated on the last 30 seconds of the 3 minutes of GM3DR+SN training data. Finally, inference was performed on all available SN-only data, which covered around 2 minutes for most volunteers. This inference was performed on very high temporal resolution. The motion model was built to represent average breathing motion, and it is therefore likely that the data acquired at this high temporal resolution will represent motion states that slightly differ from the average breathing motion states. This could decrease the correlation between the training data and the inference data, which could in turn result in a high uncertainty. To make the rejection criterion less sensitive to this effect, we evaluated it at 5 Hz rather than at 69 Hz by only flagging dynamics for which the current dynamic and all 13 preceding dynamics (acquired in around 200 ms) exceeded the rejection threshold. Note that this temporal resolution for the rejection criterion of 5 Hz is still sufficiently fast for MRgRT \citep{keall2006management,Murphy2002}. Alternatively, the SN input spokes could be averaged in time, allowing for a trade-off between SNR of the input data and the temporal resolution of the outputs. Preliminary results on this are shown in \autoref{supfig:averaging_vol3}.

We also evaluated the effectiveness of the rejection criterion based on the GP posterior uncertainty by instructing volunteer 3 and 5 in the final minutes of the SN-only phase. Volunteer 3 was instructed to perform in sequence: 1) normal breathing, 2) a switch to chest-only motion, 3) bulk motion, and 4) normal breathing. Volunteer 5 was instructed to perform 1) irregular breathing, and 2) a bulk motion of the abdomen/thorax. During these undesirable movements, we compared the rejected dynamics with another independent method based on the center-of-mass (COM). The 3D COM coordinates were computed from the 3D radial spokes \citep{anderson2013adaptive}, and the comparison was performed with the left-right and feet-head COM coordinates from a single channel of $k$-space data. 

Finally, we performed an additional analysis to gain further insight in the uncertainty estimated by the GP, and thereby the dynamics rejected by the proposed method. To this aim, we analyzed the temporal behavior of the total uncertainty $\sigma_t$ (see \autoref{sec:rejection_criterion}) over the course of the whole acquisition, which was around 10 minutes for some volunteers. A gradual build up of this uncertainty would indicate a gradual decrease in the correlation between real-time measurements and the training data, and could therefore potentially give insights in organ drifts that could affect the inference during the SN-only phase.

\section{Results}

\subsection{In silico experiments}
\begin{figure*}[tbp]
    \centering
    \includegraphics[width=.85\textwidth,height=0.28\textheight]{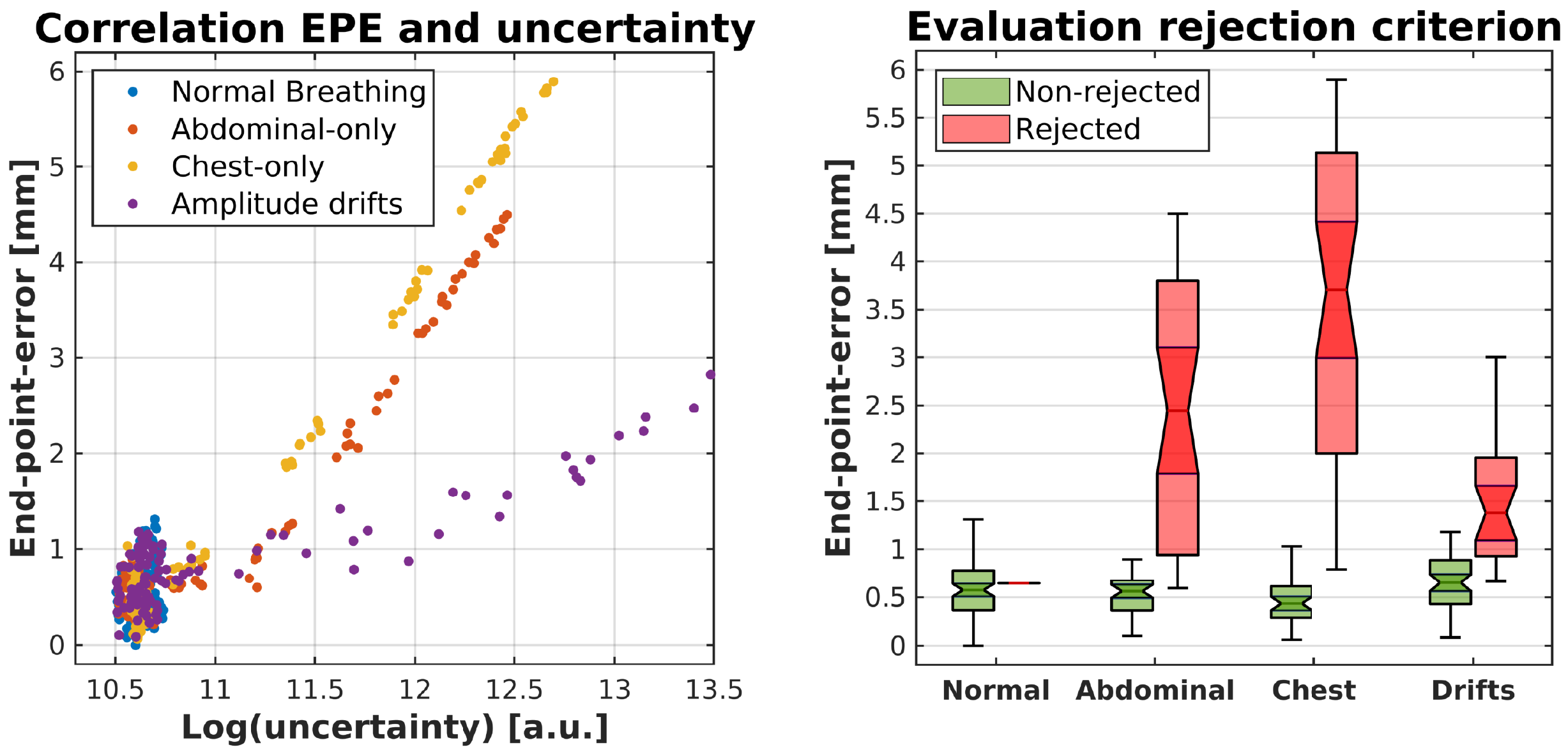}
    \caption{Left: correlation between end-point-error (EPE) and GP posterior uncertainty for all simulated breathing patterns. Note the log-scale on the $x$-axis. Additional analyses indicated for abdominal-only, chest-only and amplitude drifts, a correlation coefficient of respectively $\rho=0.99 \pm 0.00$, $\rho=0.99 \pm 0.00$ and $\rho=0.90 \pm 0.02$ $0.90 \pm 0.01$. Right: evaluation of the rejection criterion for all simulated breathing patterns; EPEs are computed before and after applying the rejection criterion. The notched boxplots indicate statistically significant ($p=0.05$) lower EPE before rejection than after rejection, showing the effectiveness to reject erroneous motion estimates.}
    \label{fig:correlation_epe_boxplots}
\end{figure*}

The left part of \autoref{fig:correlation_epe_boxplots} shows the results of the correlation analysis between GP posterior uncertainty and EPEs, indicating a positive correlation. Additionally, a statistical correlation analysis was performed between all breathing patterns other than the normal breathing pattern that was used for training. For respectively abdominal-only, chest-only and amplitude drifts, this resulted in correlation coefficients $\rho=0.99 \pm 0.00$, $\rho=0.99 \pm 0.00$ and $\rho=0.90 \pm 0.02$, which all indicate a strong positive correlation between the GP posterior uncertainty and EPEs. 

The right part of \autoref{fig:correlation_epe_boxplots} shows notched boxplots of EPEs before and after application of the rejection criterion. The results for normal breathing data, which was not exactly the same as the training data, show an interquartile range (IQR) of 0.36 mm - 0.78 mm. Here, the IQR is defined as the data between the 25$^\textrm{th}$ and 75$^\textrm{th}$ percentiles. For abdominal, chest and drifts, the IQR was respectively 0.36 mm - 0.68 mm, 0.29 mm - 0.61 mm and 0.43 mm - 0.88 mm. Overall, the maximum EPE after rejection was 1.31 mm. The notches indicate the 95\% confidence interval around the medians, and show strong statistical evidence that the EPEs of rejected dynamics are higher than non-rejected dynamics. 

Both result in \autoref{fig:correlation_epe_boxplots} empirically confirm our hypothesis that the GP posterior uncertainty can be used for real-time quality assurance, since the uncertainty is highly correlated with the error in the motion estimates. Especially the right part of \autoref{fig:correlation_epe_boxplots} highlights that the rejection criterion based on the GP posterior uncertainty rejects estimates with high EPEs and preserve low EPEs during undesirable breathing patterns. 

\subsection{In vivo experiments}

\subsubsection{In vivo robustness test}
\begin{figure*}[tbp]
    \centering
    \includegraphics[width=0.8\textwidth]{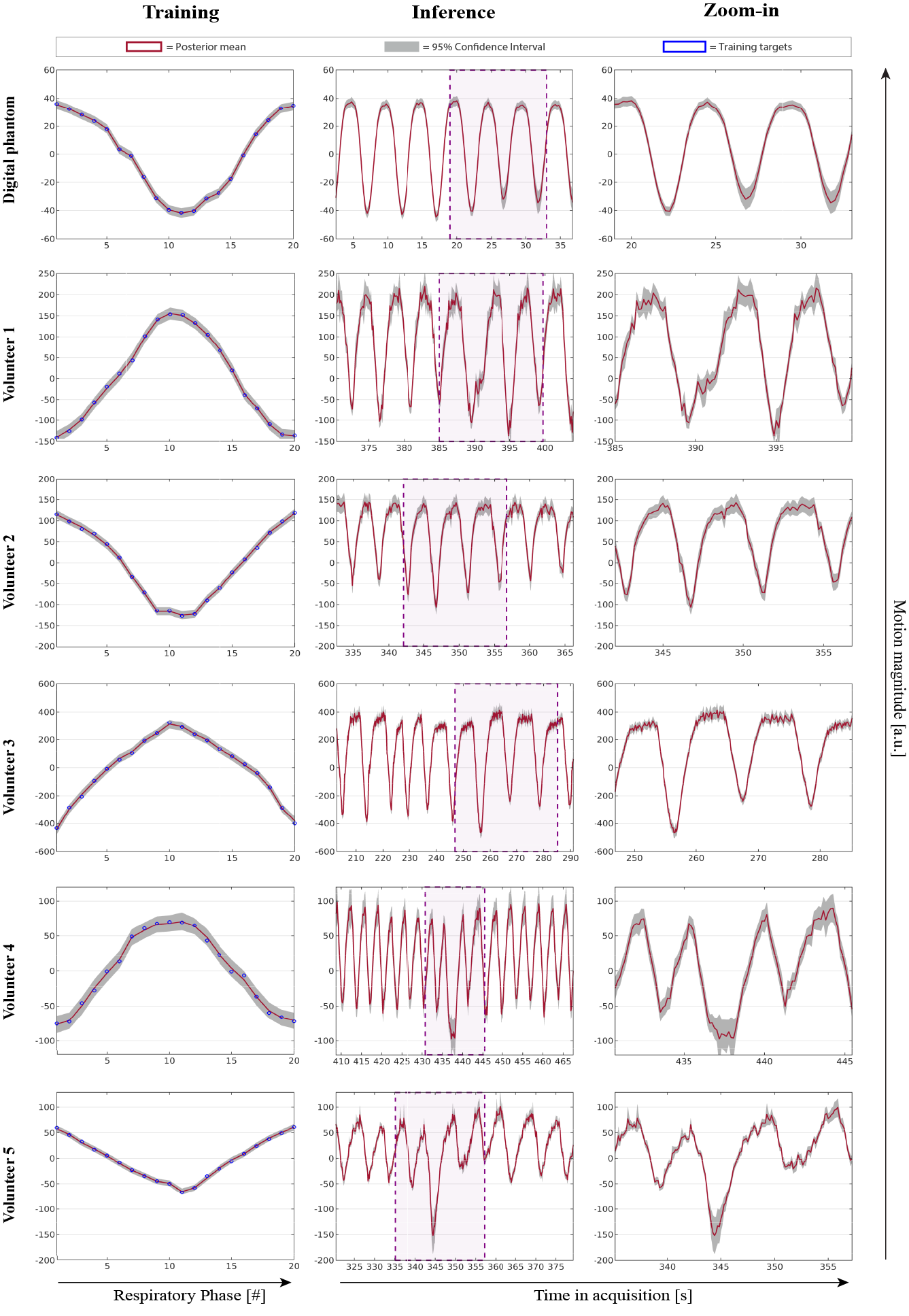}
    \caption{Qualitative results of GP motion and uncertainty estimations in a digital phantom and five volunteers. The blue dots represent the training targets, the red line the posterior mean as estimated by the first GP - which corresponds to the first principal component of the motion model - from three readouts, and the gray shaded areas are the 95\% confidence intervals obtained from the estimation uncertainty as output by the GP. The first and second columns show an overview of respectively the training and the inference phase, the third column shows a zoom-in on a region with a slightly different motion pattern or increased estimation uncertainty. The increased uncertainty mostly occurs at the highest motion amplitude levels. Note that the units on the vertical axes are arbitrary; the scaling dependents on the magnitudes of the principal components in the motion model. However, in all cases the higher values indicate exhales, while the lower values indicate inhales.}
    \label{fig:robustnesstest_allvols}
\end{figure*}

\autoref{fig:robustnesstest_allvols} shows the results of the training phase (first column), an overview of the inference phase (second column), and a zoom in a region with increased estimation uncertainty. The number of ranks in the motion model per volunteer - and thus number of GPs trained per volunteer - was determined through the L-curve approach on the explained variances, and resulted for respectively the digital phantom and volunteers 1-5 in the ranks $R=2,3,3,3,1,3,2$. The figure shows the output of the first GP, since this is generally the best interpretable component. In the first column the confidence interval should be observed; this was manually set in the training phase to account for potential errors in the training targets. The results in the overview in column 2 show realistic motion traces, and differences in breathing pattern can be observed between the different volunteers (different rows). In the zoom in column 3 it can be observed that the uncertainty increases in several scenarios: during a switch from normal to abdominal-only breathing (for the digital phantom, $t\approx25$ s), during deep inhales (volunteers 4 and 5) and during deep exhales (volunteers 1 and 2). The increased uncertainty is especially pronounced for dynamics with breathing amplitudes outside the range of the training data, i.e. larger than average exhales or inhales. This observation is explainable, since the GP posterior uncertainty increases with the distance to the training data, as evaluated by the kernel function. Animated Figure 1 in the supporting files shows the posterior mean and spatial estimation uncertainty maps for volunteer 1, as derived in \eqref{eq:gp_dvf_distribution}. The animated figure shows the inference over the first 35 seconds in the second column in \autoref{fig:robustnesstest_allvols}, and visualizes every 4$^\textrm{th}$ dynamic with a total 60 frames at 4 Hz. Similar to the results in \autoref{fig:robustnesstest_allvols}, the respiratory traces appear smooth overall, however small high frequency oscillations can be observed which could be related to cardiac motion or measurement imperfections (eddy currents). We refer the reader to \ref{section:appendix} for an overview of all animated figures.

\begin{figure*}[tbp]
    \centering
    \includegraphics[width=\textwidth]{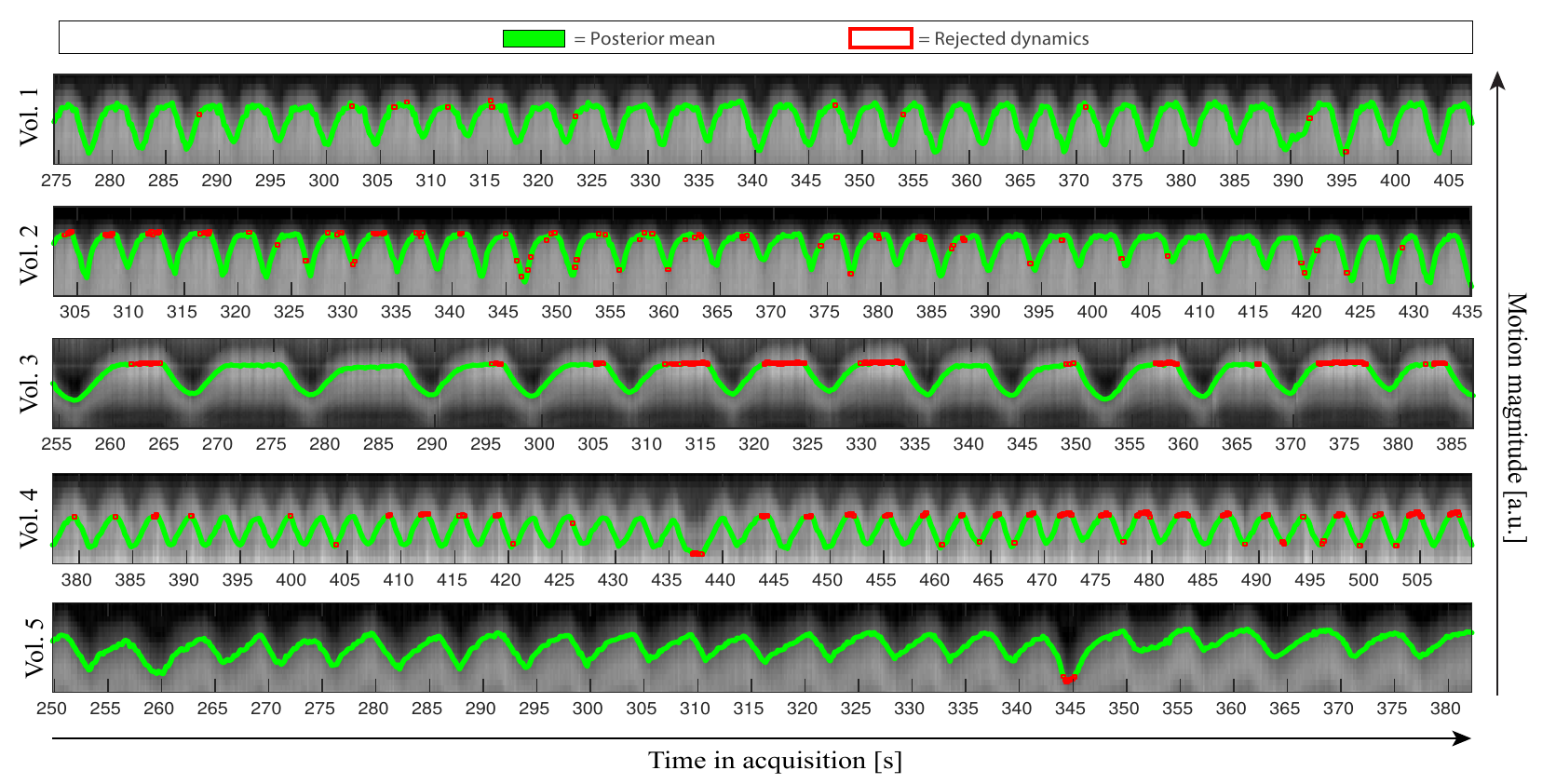}
    \caption{Comparison between GP predictions and projection images. The image shows the comparison for all five volunteers between the posterior mean as estimated by the first GP corresponding to the first principal component of the motion model (green), projections on the FH-axis as obtained by an FFT over the readout of FH spokes (background), and rejected dynamics (red marks). The inference was performed on data acquired about two minutes after the training data were acquired.}
    \label{fig:robustness_projections_gm3dr}
\end{figure*}

\begin{figure*}[tbp]
    \centering
    \includegraphics[width=.9\textwidth]{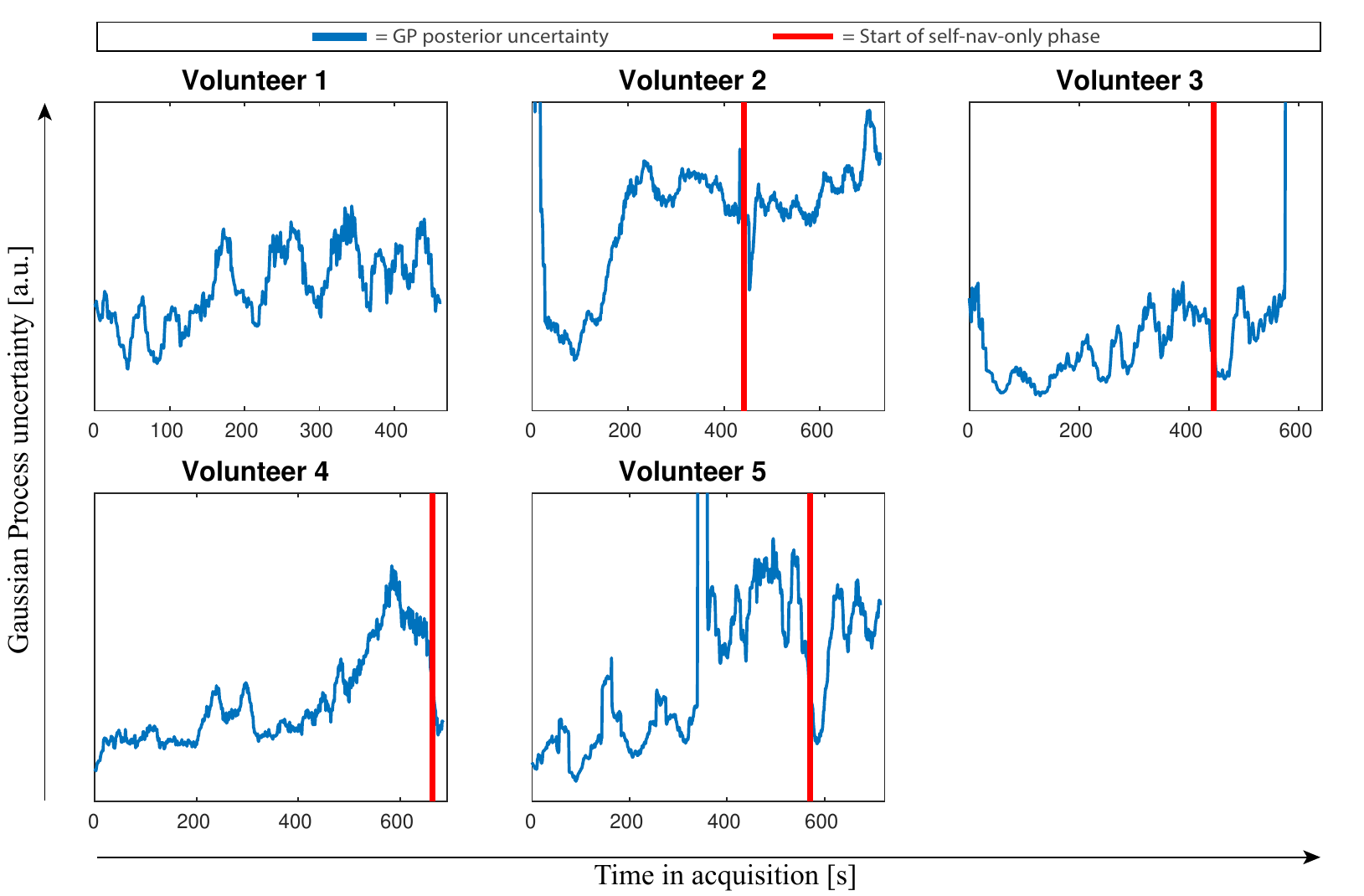}
    \caption{Analysis of the GP posterior uncertainties over the course of the whole acquisition. For volunteers 2-5 the acquisition switched to an SN-only acquisition at the end of the acquisition, indicated by the solid red line. A moving average filter with a window of 10 seconds is applied to the uncertainties to highlight the patterns over the whole acquisition rather than within the breathing cycles.}
    \label{fig:drift_analysis_allvols}
\end{figure*}

The validation by comparison with the projection images obtained from FH $k$-space spokes is shown in \autoref{fig:robustness_projections_gm3dr}. Moreover, the figure shows the results of the application of the rejection criterion, as discussed in \autoref{sec:rejection_criterion}. Overall the posterior mean of the first GP coincides with the projection images. The framework mostly rejected end-exhale dynamics, which were likely outside the range of the training data. Interestingly, for volunteer 2 the frequency of rejections is relatively large at the beginning, and decreases after about 400 seconds in the acquisition. For volunteers 4 and 5, it can be observed that the framework also rejected deep inhales, which were substantially different from the average breathing in the training data. The change in the frequency of rejections could be an indication of a breathing pattern change. The large number of rejections in exhale could be due to drifts of internal organs, resulting in a different end-exhale position. These hypotheses are further analyzed in \autoref{fig:drift_analysis_allvols}.

\subsubsection{Feasibility tests of high-speed inference at 69 Hz}
\label{section:highspeedfeasibility_results}
\begin{figure*}[tbp]
    \centering
    \includegraphics[width=\textwidth]{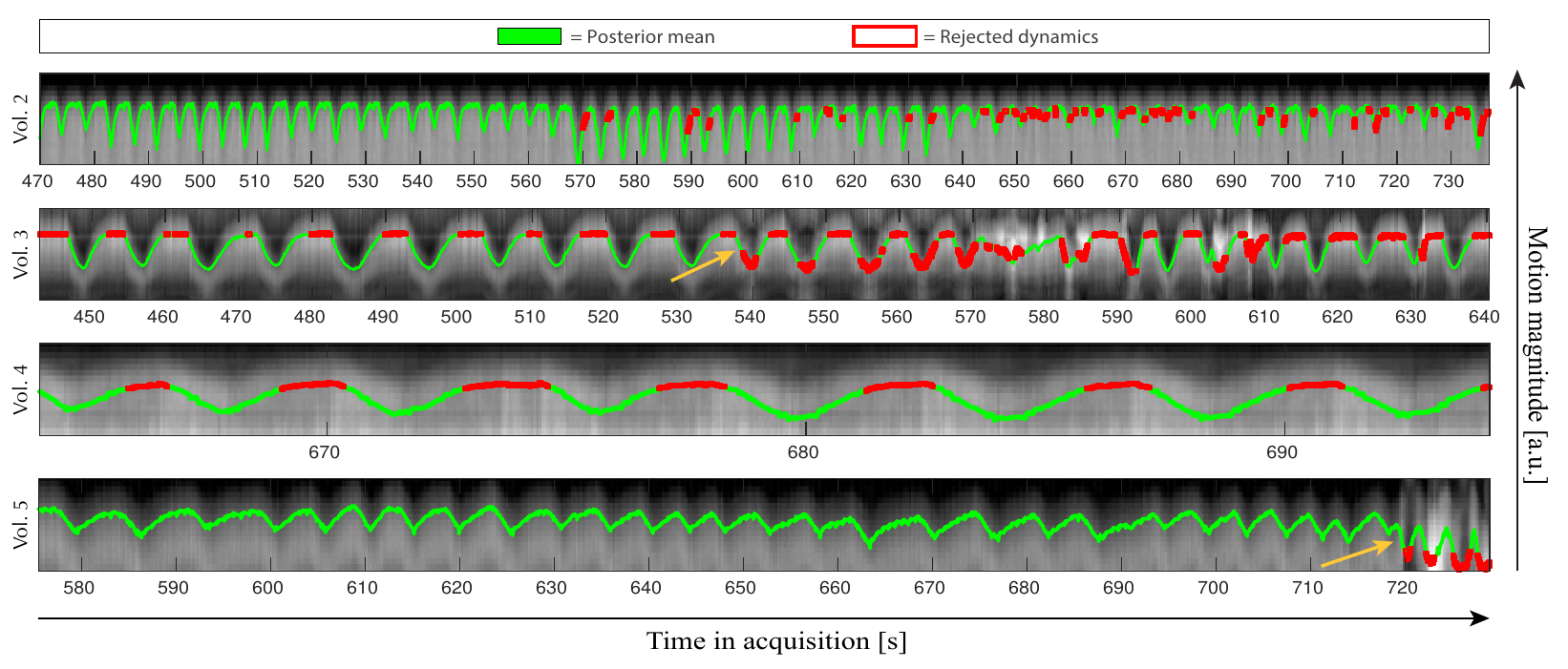}
    \caption{High-speed in-vivo inference and quality assurance. The posterior mean of the first GP (green), and rejection criterion evaluations (red), are compared with projections on the FH-axis for all four volunteers with SN-only data available (2-5). Volunteers 3 and 5 were instructed to perform motion which would render the motion model invalid. Volunteer 3 performed in sequence: 1) normal breathing; 2) a switch to chest-only breathing; 3) bulk motion; 4) normal breathing, and the first abnormal event started around 535 seconds in the acquisition, as indicated by the yellow arrow. Volunteer 5 performed in sequence: 1) normal breathing; 2) irregular breathing; 3) bulk motion, and the first abnormal event started around 718 seconds in the acquisition, as indicated by the yellow arrow. The GP posterior means are scaled to visually overlap the pattern in the projection images.}
    \label{fig:highspeed_inference_allvols}
\end{figure*}

In this experiment, inference was performed at 69 Hz, using only three mutually orthogonal readouts as input. The results are shown in \autoref{fig:highspeed_inference_allvols}, which compares the GP inference with projection images (background). The Animated Figures 2-5 in the supporting files show the posterior mean and spatial estimation uncertainty maps for volunteers 2-5, as derived in \eqref{eq:gp_dvf_distribution}. These animated figures show the inference over the first 35 seconds of the data in \autoref{fig:highspeed_inference_allvols}. The animation shows every 40$^\textrm{th}$ dynamic with a total 60 frames, visualized at 4 Hz. An exception to this is volunteer 4, for whom every 80$^\textrm{th}$ dynamic is visualized at 4 Hz because of a low breathing frequency. We refer the reader to \ref{section:appendix} for an overview of all animated figures.

Overall the inference during normal breathing returns plausible motion traces. Volunteer 2 shows a regular breathing pattern, which changes to larger breathing amplitudes halfway the scan, and eventually shallow breathing at the very end of the scan. Volunteer 3 and 4 show a regular breathing pattern, and volunteer 5 shows a very irregular breathing pattern. The irregularities in the breathing are mostly accompanied by larger uncertainty, which is most pronounced for motion states that fall outside the range of motion in the training phase. The spatial uncertainty map for volunteer 3 shows a larger uncertainty than the other volunteers, although the visualized uncertainty range of 0 mm - 1 mm should be noted here. The relatively larger uncertainty could be related to the relatively large breathing amplitudes of volunteer 3.

For volunteer 2, hardly any dynamics were rejected until about 640 seconds in the acquisition, after which the breathing pattern seems to change to very shallow breathing, resulting in many rejections. For volunteers 3 and 5, hardly any dynamics were rejected before the instructed motion changes, which started at respectively 535 seconds and 718 seconds in the acquisition. For volunteer 4, almost all end-exhale dynamics were rejected. Since this occurs after more than 10 minutes in the acquisition, this could be caused by organ drifts that changes the internal positions of the organs in such a way that it leads to rejections. This hypothesis is further analyzed in \autoref{fig:drift_analysis_allvols}. The results show that for most volunteers the posterior uncertainty gradually increases over the course of the acquisition, which indicates a gradual decrease of the correlation between the acquired data and the data in the training set. This could be explained by physiological drifts, which are also visually observable in \autoref{fig:robustnesstest_allvols} and the projection images over the course of the whole scan (not shown). 

\begin{figure*}[tbp]
    \centering
    \includegraphics[width=\textwidth]{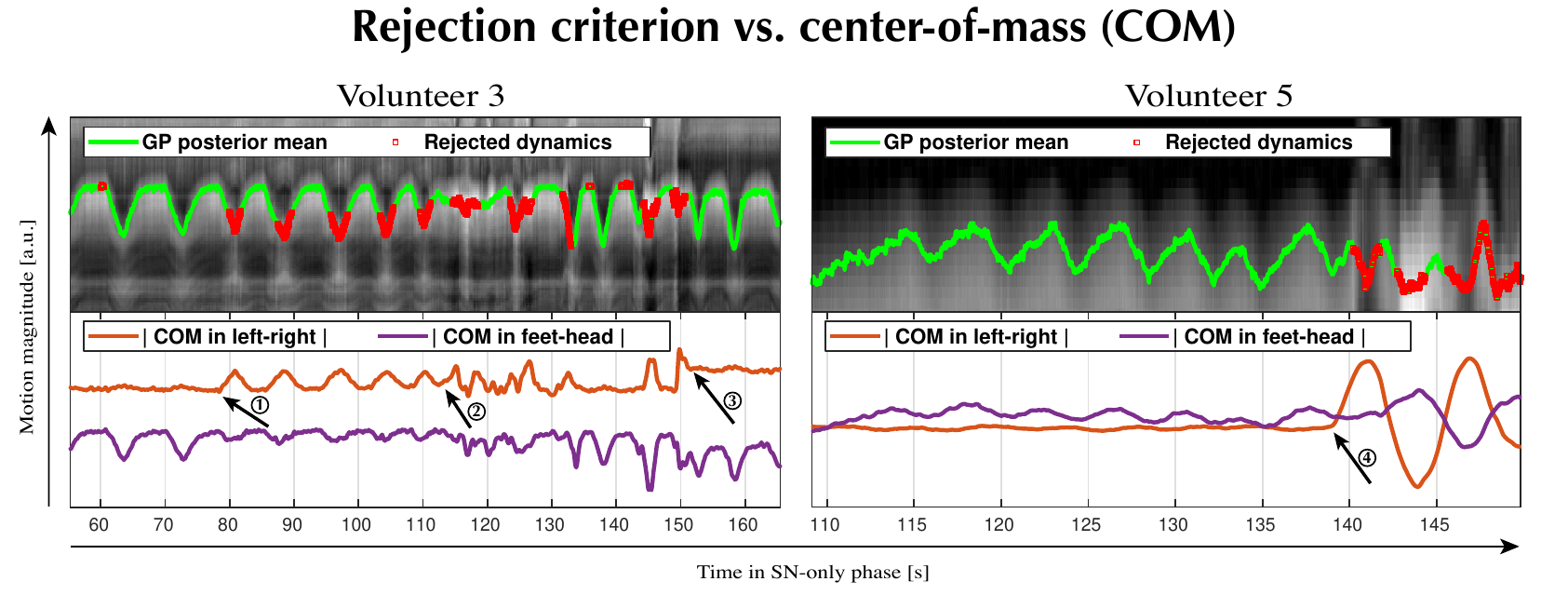}
    \caption{This figure shows the result of the high-speed inference for volunteers 3 (left) and 5 (right), as discussed in \autoref{section:highspeedfeasibility_experiments} and \autoref{section:highspeedfeasibility_results}. The bottom graphs show a comparison with center-of-mass (COM) coordinates in left-right (orange) and feet-head (purple). The black arrows, labeled 1-4, indicate four events where the motion was changed, as can also be observed from the changing patterns in the COM-coordinates; 1) volunteer 3 switches breathing pattern, 2) volunteer 3 performs bulk motion, 3) volunteer 3 returns to normal breathing, 4) volunteer 5 performs bulk motion.}
    \label{fig:robustness_projections_sn}
\end{figure*}

To evaluate the rejection criterion, volunteer 3 and 5 were asked to perform a specific motion that forced a deviation from the data in the training phase and would render the motion model inapplicable. The results are shown in \autoref{fig:robustness_projections_sn}. The top figure shows a comparison with the projection images obtained with an FFT along the readout of the FH $k$-space spokes, as well as the evaluation of the rejection criterion, as discussed in \autoref{sec:rejection_criterion}. The bottom part of the figure shows the center-of-mass (COM) coordinates in left-right (orange) and feet-head (purple), extracted from the 4$^\textrm{th}$ channel of $k$-space data. The COM coordinates serve as an independent visualization of the volunteer's behavior during this experiment. The black arrows indicate the changes in motion, which are accompanied by a change in the pattern of the COM. Volunteer 3 (left), changed breathing pattern at arrow \#1, performed bulk motion at arrow \#2, and returned to a normal breathing pattern at arrow \#3. Volunteer 5 shows irregular breathing with a slight drift towards inhale up to arrow \#4, and bulk motion after arrow \#4. The framework successfully detected abnormal motion, and rejected many dynamics. It should be noted that Volunteer 3 appears to return to a different position after the bulk motion, since the COM in left-right is slightly higher in the final part of the figure. Although this is not reflected by the framework's rejections, in practice the large number of consecutive rejections preceding this event should already be indicative that the model's predictions are too uncertain, and that the motion model should thus be updated. For volunteer 5, almost all dynamics were rejected during bulk motion. Only two sets of consecutive dynamics were not rejected during bulk motion, during which the COM coordinates indicate a similar position as in the preceding normal breathing phase.

\section{Discussion}
\subsection{Summary of innovations}
In this work we presented a probabilistic framework which addresses two major technical hurdles towards real-time adaptive MRgRT simultaneously: real-time 3D motion-field estimation and uncertainty quantification. The framework was built on the idea that low-dimensional motion information can be extracted from few readouts of $k$-space data \cite{huttinga2021realtime}. This idea was exploited via a two-step reconstruction approach, in which first a motion model was built, and subsequently the model's coefficients were inferred from the data. 

For the inference, a probabilistic machine learning regression technique based on Gaussian Processes was used. Due to its probabilistic nature, this technique not only estimated the most likely motion model coefficients, but also provided a measure of estimation uncertainty. The inferred model coefficients combined with the motion model yielded the motion-fields required for MRgRT. The inferred uncertainty was hypothesized to be useful for real-time quality assurance during radiotherapy. This hypothesis was empirically confirmed in simulations, in which it was shown to enable the detection of erroneous motion estimates, which - if left undetected - could in practice result in harmful radiation to organs-at-risk.

As opposed to other machine learning techniques such as neural networks, the training of the GPs could be done in about half a second, and required only a minimal amount of training samples (20 in this work). Moreover, the inference time was very short ($\approx 0.1$ milliseconds per dynamic) due to the availability of a rapidly computable closed-form analytical expression for the GP's posterior distribution. Altogether, this makes the GP a natural fit in a real-time MRgRT workflow where little latency is essential. 

The complete framework was extensively validated in silico with simulations using a digital phantom, and in vivo with MR-linac data of five volunteers, thereby taking into account different breathing patterns and bulk motions.

\subsection{Potential impact}

The presented framework can have several applications in MR-guided radiotherapy. Firstly, the framework allows to infer 3D motion-fields at 69 Hz. The inference at this speed is more than sufficient to resolve abdominothoracic motion during MR-guided radiotherapy \citep{keall2006management,Murphy2002}. Although 5 Hz should be sufficient for this application, the feasibility of inference at 69 Hz could open up possibilities of applying radiotherapy to tumors subject to cardiac motion, such as central lung tumors. Another application could be cardiac radio-ablation \citep{cuculich2017noninvasive}. This is an emerging non-invasive treatment technique of cardiac arrhythmias with highly focused radiotherapy, which could benefit from high-speed tracking of myocardial landmarks.

Secondly, the proposed framework not only estimates motion-fields, but also provides a measure of estimation confidence. We have demonstrated an example of how this measure of confidence can be used for real-time quality assurance by designing a rejection criterion based on this estimation confidence. In practice, this could be useful to detect unexpected motions during radiotherapy, such as a change of breathing pattern or bulk motions, which could result in erroneous motion estimates due to an unsuitable motion model built for normal breathing. Without any measure of quality assurance, such estimates could result in harmful radiation to organs-at-risk. With the real-time quality assurance proposed in this work, the treatments could be (temporarily) halted to assure the patient's safety during such potentially erroneous motion estimates.

Finally, the availability of time-resolved 3D motion-fields over the course of a radiotherapy treatment could be used for retrospective dose accumulation calculations \citep{Kontaxis2020}. Such calculations provide insights in the actual dose deposited to the target tumor and surrounding organs-at-risk during a treatment, which can be taken into account to improve subsequent treatment planning.

\subsection{Related work}
Real-time 3D inference as proposed in this work was shown before. For example, our own method MR-MOTUS \citep{huttinga2021realtime} achieved 3D motion-field reconstruction at 6.7 Hz, \citep{li2010real} reconstructed 3D CT volumes and motion-fields at about 4 Hz, MR-SIGMA \citep{Feng2020} estimated 3D MRI volumes at 3.3 Hz, and cine-based MRI methods such as proposed in \citet{stemkens2016image} achieved 3D motion-field reconstruction at about 2 Hz. Slightly different type of methods are based on surrogate signal models \citep{Mcclelland2017,mcclelland2013respiratory,andreychenko2017thermal,Low2005,Tran2020}, and also inferred 3D motion-fields at high temporal resolution from (several) 1D surrogate signals. Recently, also several deep learning (DL) based methods were proposed for real-time 3D inference. For example, \citet{ROMAGUERA2021102250} inferred 3D motion-fields from a pre-treatment volume and real-time 2D images, and also predicted the motion for future timepoints. \citet{terpstra2021real} proposed TEMPEST, a network that estimates 3D motion-fields directly from highly undersampled, aliased, 3D images with a frame rate up to 5 Hz. 

Frequently, non-DL-based real-time inference methods exploit a low-rank motion model similar to the one employed in this work, which is typically obtained by a retrospective compression of motion-fields using principal component analysis \citep{zhang2007patient,stemkens2016image,mishra2014initial,cai20153d,king2012thoracic}, or by decoupling the motion-fields into spatial components and temporal components based on surrogate signals \citep{Low2005,mcclelland2013respiratory,Mcclelland2017}. Our previous works \citep{Huttinga2021,huttinga2021realtime} also employ a low-rank model, but estimate its components directly from $k$-space data by solving a large-scale non-linear optimization problem \citep{huttinga2020mr}. 

Most of the methods above only estimate motion-fields, without any measure of confidence or motion model applicability. An exception is the work by \citet{king2012thoracic}, where a low-rank 3D motion model was fit to incoming 2D navigator MR-images and the motion model's applicability was constantly evaluated as the image similarity after registration. A notable difference is that \citet{king2012thoracic} used 2D cine navigator images, whereas we used three 1D spokes. The network proposed by \citet{ROMAGUERA2021102250} also outputs estimation uncertainties, but these were not used for quality assurance. The proposed rejection criterion is therefore most similar to the model applicability test in \citet{king2012thoracic}. 

The employed dimension reduction techniques for both the input and output space were proposed to make GPs fit in a high-dimensional pipeline. Similar dimension reduction techniques on the output space were previously proposed in PCA-GPs \citep{higdon2008computer}, with the aim to overcome the same challenges of combining GPs with high-dimensional output data. In this work, GPs were used to simultaneously perform regression over multiple scalar functions with multi-dimensional inputs, thereby assuming no correlations between the individual GP outputs. Alternatively, correlation in the outputs could be modeled with multi-task GPs \citep{bonilla2007multi}. Although this could improve the performance in theory, our preliminary results in \citep{sbrizzi2019acquisition} with multi-task GPs showed little to no improvement.

\begin{figure*}[h!]
    \centering
    \includegraphics[width=.95\textwidth]{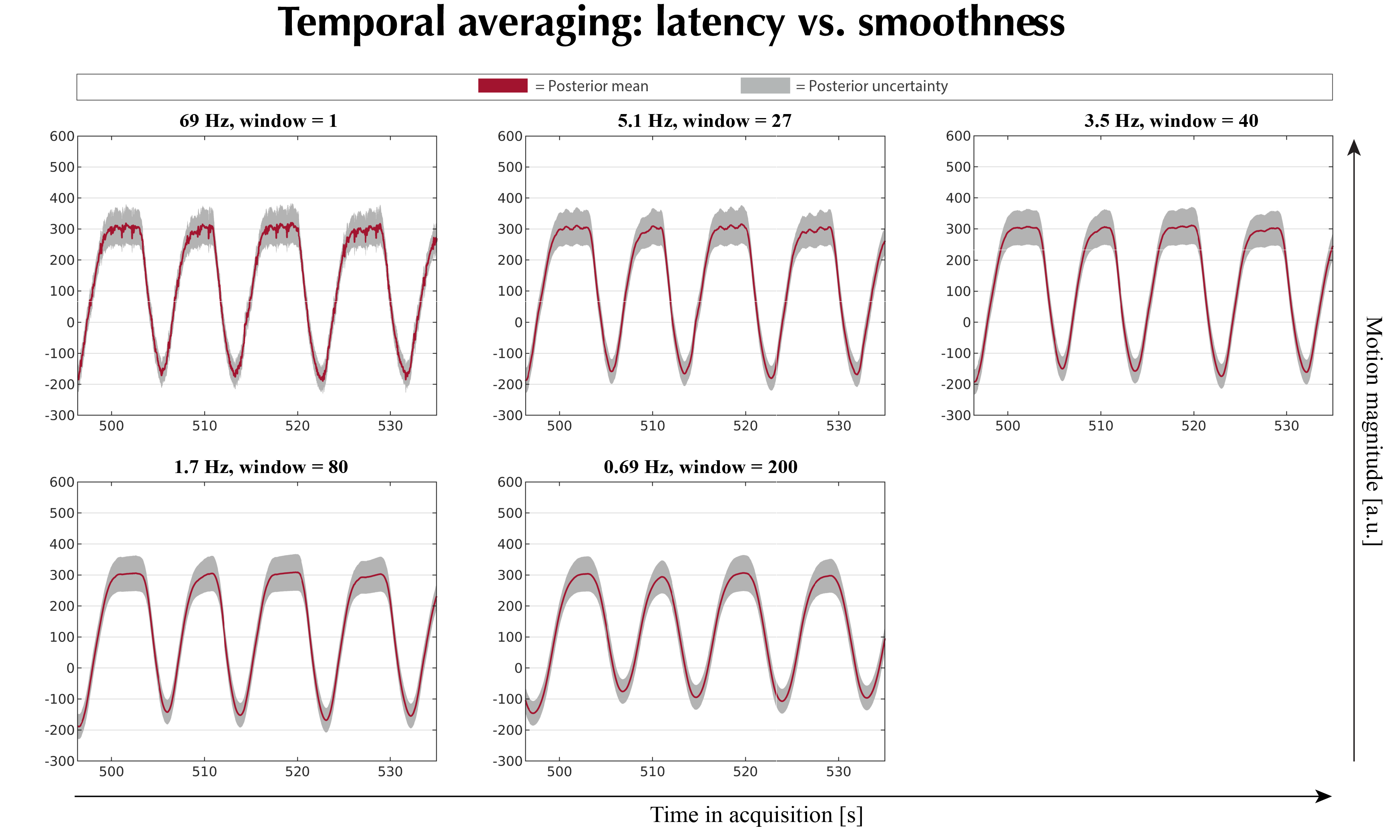}
    \caption{The effect of temporally averaging the input SN-spokes. Several reconstructions of the first GP for volunteer 3 are shown for which a symmetric moving mean filter is applied with a varying width: 1, 27, 40, 60, 80, and 200 dynamics. This averaging increases the temporal latency, as indicated in the titles, since the data required by the second half of the window lies in the future. However, evidently, the filtering also results in smoother reconstructions. Hence, temporal filtering allows to make a trade-off between the temporal resolution of the reconstructions and the SNR. Temporal averaging with a width of 27 dynamics would result in sufficient temporal resolution for real-time adaptive MRgRT \citep{Murphy2002,keall2006management}.}
    \label{supfig:averaging_vol3}
\end{figure*}

\subsection{Points of improvement and future work}
Several aspects of this work could be improved for a clinical application. The proposed GP framework provides all the utilities required for real-time adaptive MRgRT. Further improvements on the method are modular and not strictly necessary. We expect the largest gains in performance to reside in improvements to the motion model, for example by better image reconstructions, or better image registrations. The number of rejected dynamics could possibly be reduced whenever a more expressive motion model would be used, e.g. a model built on 3D+t time-resolved cine MR-images, as proposed in \citet{king2012thoracic}. 

We have extensively validated our framework in silico with end-point-errors and four different breathing patterns. Moreover, the performance was assessed in vivo by comparisons with FH projection images. For a practical application, more in vivo validations are required with e.g. manually tracked targets on cine MR-images. This will be subject of a future work.

Several volunteers indicate a sensitivity of the framework to high-frequency oscillations in the data. These oscillations could be physiological, e.g. due to local cardiac motion or related blood flows. In the current framework, these high-frequency oscillations could not be coupled to local motion since this was not incorporated in the motion model. More complex motion models could possibly allow to identify the source of these oscillations. Alternatively, since inference could be performed with a speed of 69 Hz - which is about 10 times the required speed for MRgRT - the oscillations could simply be filtered out by performing a temporal averaging on the GP inputs. The averaging would allow for a trade-off between temporal resolution and smoothed outputs. Preliminary results in \autoref{supfig:averaging_vol3} show that this is indeed feasible. 

In this work, only correlations in the data were considered, but temporal correlations were not yet taken into account. Such an extension would require to model temporal correlations, but could prove valuable for near-future predictions. A possible downside of including temporal correlations is that it could impose strong restrictions on the temporal behavior of the motion (e.g. periodicity), which is why we have not considered it in this work. 

In general, it should be noted that all improvements discussed above are modular, and would require little to no changes in the general pipeline of the proposed framework.

\section{Conclusion}
We have presented a probabilistic framework for simultaneous real-time 3D motion and uncertainty estimation. The complete framework, including the rejection criterion, allowed to preserve low EPEs ($75^\textrm{th}$ percentiles $\le 0.88 $ mm) during four different breathing patterns in simulations. Without the proposed rejection criterion these breathing patterns would have resulted in EPEs up to almost 6 mm, which - if left undetected - could lead to harmful radiation to organs-at-risk. The framework estimated in vivo motion that corresponds well with FH projections. Moreover, it flagged dynamics during which bulk motion and changes of breathing patterns were performed. This flagging strategy could be used to ensure the safety of the patient by (temporarily) halting the radiation, and continuing whenever confidence in the estimated motion is restored. Altogether, the framework tackles two major technical challenges for real-time adaptive MR-guided radiotherapy, real-time 3D MR-based motion estimation and uncertainty quantification, and it thereby paves the way to reach the ultimate potential of the MR-linac.

\appendix

\section{Animated figures}
\label{section:appendix}
All supporting animated figures corresponding to this manuscript are available at \url{https://surfdrive.surf.nl/files/index.php/s/scLts9nJYXfbLMx}. 

\subsection*{Animated Figure 1}
This GIF shows the posterior mean and spatial estimation uncertainty maps for volunteer 1, as derived in \autoref{eq:gp_dvf_distribution}. The animated figure shows the inference over the first 35 seconds in the second column in \autoref{fig:robustnesstest_allvols}, and visualizes every 4$^\textrm{th}$ dynamic with a total of 60 frames at 4 Hz.

\subsection*{Animated Figures 2-5}
These GIFs show the posterior mean and spatial estimation uncertainty maps for volunteers 2-5, as derived in \autoref{eq:gp_dvf_distribution}. These animated figures show the inference over the first 35 seconds of the data in \autoref{fig:highspeed_inference_allvols}. The animation shows every 40$^\textrm{th}$ dynamic with a total of 60 frames, visualized at 4 Hz. An exception to this is volunteer 4, for whom every 80$^\textrm{th}$ dynamic is visualized at 4 Hz because of a low breathing frequency.

\section*{Acknowledgements}
This work was supported in part by the Dutch Research Council (NWO) under Grant 15115.

\bibliographystyle{latex/model2-names.bst}
\biboptions{authoryear,sort}
\bibliography{references}

\end{document}